\documentclass[
 amsmath,
 amssymb,twocolumn,
 aps,prx]{revtex4-2}

\usepackage{graphicx}
\usepackage{dcolumn}
\usepackage{bm}
\usepackage{hyperref}
\usepackage{float}
\usepackage[normalem]{ulem}
\DeclareUnicodeCharacter{2212}{-}

\usepackage{color}
\usepackage[dvipsnames]{xcolor}

\begin{document}

\preprint{APS/123-QED}

\title{Squeezing as a resource for time series processing in quantum reservoir computing}

\author{Jorge Garc{\'\i}a-Beni}
\author{Gian Luca Giorgi}
\author{Miguel C. Soriano}
\author{Roberta Zambrini}

\affiliation{
 Instituto de F{\'\i}sica Interdisciplinar y Sistemas Complejos (IFISC), UIB–CSIC \\
 UIB Campus, Palma de Mallorca, E-07122, Spain 
}

\date{\today}
\begin{abstract}
Squeezing is known to be a quantum resource in many applications in metrology, cryptography, and computing, being related to entanglement in multimode settings.
In this work, we address the effects of squeezing in neuromorphic machine learning for time series processing. In particular, we consider a loop-based photonic architecture for reservoir computing and address the effect of squeezing in the reservoir, considering a Hamiltonian with both active and passive coupling terms. 
 Interestingly, squeezing can be either detrimental or beneficial for quantum reservoir computing when moving from ideal to realistic models, accounting for experimental noise. 
We demonstrate that multimode squeezing enhances its accessible memory, which improves the performance in several benchmark temporal tasks. The origin of this improvement is traced back to the robustness of the reservoir to readout noise as squeezing increases.
\end{abstract}

\maketitle

\section{Introduction} \label{Intro}
Squeezing is a quantum phenomenon characterized by reduced light field quadrature fluctuations below shot noise levels \cite{ferraro2005gaussian,adesso2014,serafini2017}. Initially employed in fundamental quantum tests, such as Einstein–Podolsky–Rosen (EPR) paradox experiments \cite{EPR1,EPR2}, squeezing has emerged as a crucial resource in diverse quantum technologies. Notably, squeezed states have been extensively utilized in quantum metrology to enhance measurement sensitivity for parameter estimation \cite{Giovannetti2004,Berni2015}, clock synchronization \cite{Giovannetti2001}, and gravitational wave detection \cite{LIGO1}. Moreover, their role as a resource for quantum entanglement has been harnessed for quantum cryptography protocols \cite{Madsen2012,Gehring2015}. In boson sampling experiments, large multimode squeezed states have made it possible to achieve a quantum advantage \cite{Zhong2020-boson,Madsen2022}. Additionally, they serve as the primary resource for universal measurement-based quantum computing in continuous variables (CV) \cite{universal-clusters} through the generation of cluster states \cite{furusawa-big-cluster,60-mode-frequency,furusawa-2D-cluster}. 
In the context of quantum machine learning, squeezing is fundamental for CV quantum neural networks to outperform their classical counterparts \cite{PhysRevResearch.1.033063}.
In this work, we will focus on the favorable impact of squeezing on time-series prediction and forecasting in the context of Quantum Reservoir Computing (QRC) \cite{Mujal_opportunities}.

Reservoir Computing (RC) constitutes an unconventional paradigm within the realm of machine learning techniques rooted in recurrent neural networks \cite{jaeger2001echo,maass2002real,LUKOSEVICIUS2009127}. Particularly tailored for time series processing, RC allows fast learning with minimal training costs. The RC framework has demonstrated its effectiveness in real-world scenarios including temporal prediction tasks \cite{WYFFELS20101958,LIN20097313,Ilies07steppingforward,RCbook} as well as classification tasks \cite{NIPS2010_2ca65f58,WANG2016237}. By harnessing the information processing capabilities of high-dimensional dynamical systems, RC concepts have seamlessly transitioned to physical substrates \cite{tanaka2019}, with photonic and optoelectronic implementations receiving attention for their high-speed attributes \cite{brunner2013parallel,vandoorne2014experimental,larger2017high,VanDerSande2017}.

Recently, the scope of RC has expanded to encompass quantum systems, capitalizing on their augmented Hilbert space for enhanced performance \cite{Mujal_opportunities}. Notably, quantum enhancements in temporal tasks under ideal conditions have been observed in both spin \cite{Fujii-Nakajima,rodri-IPC} and photonic setups \cite{Nokkala2021,qu-memristor}. 
Different aspects influencing  quantum reservoirs perfomance have been considered and the effectiveness of complex task solving has been addressed considering different evolution maps \cite{Fujii-Nakajima,chen-nurdin,sannia2022dissipation,qu-memristor}, the role of statistics \cite{Qproc,guillem_particle}, or different quantum phases \cite{rodrigo_PRL}.
These investigations assumed ideal conditions, attributing performance improvements to factors such as improved memory properties, more favorable nonlinearities, and expanded accessible Hilbert spaces.

However, QRC faces several challenges, with substantial attention directed towards addressing the presence of noise in output observables \cite{Johannes_fluctuations,time-series-QRC-measurements,garciabeni2023,hu2023tackling,Kalfus2022}. While readout noise is relevant to classical RC as well \cite{Soriano_noise_1,Soriano_noise_2}, it acquires heightened significance in QRC due to intrinsic sampling noise arising from the stochastic nature of quantum measurements. This noise significantly hinders potential quantum enhancements \cite{time-series-QRC-measurements,garciabeni2023,hu2023tackling}. 
The strategy of monitoring the output accounting on the effects of quantum measurement for temporal tasks is also particularly critical \cite{Mujal2023,garciabeni2023}.
Preserving quantum advantage within these non-ideal circumstances is needed to ensure the viability of QRC protocols.

 In Ref. \cite{Mujal2023} it was shown how weak --instead of projective-- measurements allow continuous monitoring in QRC --instead of buffering inputs and rewind the reservoir dynamics. Continuous homodyne monitoring in QRC was addressed in Ref.  \cite{garciabeni2023} in a photonic platform. Here,  a loop-based photonic platform was proposed, suitable for online time series processing. These works show the possibility of sustaining the RC performance in the quantum setting in the presence of non-ideal measurement conditions. Motivated by these results,  in this work, we adopt a slightly simplified version of the photonic platform used in Ref. \cite{garciabeni2023} and investigate how the presence of squeezing in the optical cavity 
can improve the performance of the reservoir. We will numerically analyze the role of squeezing, addressing active and passive coupling terms in a photonic network, in both linear and nonlinear tasks. An analytical argument is made to justify such an improvement.

The paper is structured as follows: in Sect. \ref{sec-II}, the general framework of RC as well as a detailed description of our photonic platform are exposed. In Sect. \ref{sec-III} the simulation results for some benchmark RC tasks under non-ideal noise conditions are shown: we test the linear memory of the system (Sect. \ref{sec-IIIA}) as well as the non-linear memory (Sect. \ref{sec-IIIB}) and its performance on time-series forecasting (\ref{sec-IIIC}). In Sect. \ref{sec-IV} we show numerical evidence to explain the noise robustness improvement caused by squeezing in the previous tasks. Finally, conclusions are given in Sect. \ref{conclusions}.
\section{Loop-based architecture} \label{sec-II}
\subsection{Reservoir computing} \label{RC-section}
RC architectures are mainly composed of three distinct layers: the input layer, the reservoir, and the readout layer \cite{ingo-nakajima}. The external signal is encoded and fed into the system in the input layer. This is done sequentially at each time step. The reservoir layer (or just reservoir) is usually a complex dynamical system that applies a non-linear map to the inputs. The reservoir must retain short-term memory of previous inputs to be able to perform temporal tasks. This short-term or fading memory, together with the echo state property, is part of the universality proofs of RC \cite{GRIGORYEVA2018495}. The readout layer is then made of a certain number of reservoir observables, which are monitored sequentially after the input is encoded. The output from this layer is a linear combination of the measured observables. Supervised learning is performed by optimizing this linear combination of the output to yield the desired target. 

In more detail, if we have a training set consisting of a sequence of $L$ inputs $\left\{ s_{1}, s_{2}, \dots, s_{L} \right\}$, at time step $k$ the input $s_{k}$ is encoded and introduced into the reservoir. If we call $\mathbf{x}_{k}$ to the reservoir degrees of freedom, we can write the reservoir map at time step $k$ as 
\begin{equation}
\mathbf{x}_{k} = \mathcal{H}\left(\mathbf{x}_{k-1}, s_{k}\right) .
\end{equation}
This map is fixed throughout the whole protocol. For the readout layer at time step $k$, we use $\mathbf{O}_{k}$ as readout observables (functions of $\mathbf{x}_{k}$). The reproduced function at each time is obtained by performing a linear regression on the readout observables, 
\begin{equation} \label{y-linear-reg}
    y_{k} = w_{0} + \mathbf{O}_{k}^{\top} \mathbf{w} \\
    = \left( 1 , \mathbf{O}_{k}^{\top} \right) \left( \begin{array}{c}
        w_{0}  \\
        \mathbf{w}
    \end{array} \right) ,
\end{equation}
where the weight vector $\mathbf{W} = \left( w_{0}, \mathbf{w}^{\top} \right)^{\top}$ is optimized through training examples. The way training works is as follows: for the sequence of $L$ training inputs, we define the matrices
\begin{equation}
    V = \left(
    \begin{array}{cc}
        1 & \mathbf{O}_{1}^{\top} \\
        1 & \mathbf{O}_{2}^{\top} \\
        \vdots & \vdots \\
        1 & \mathbf{O}_{L}^{\top}
    \end{array}
    \right)  ; \quad  \mathbf{y} = \left( \begin{array}{c}
        y_{1}  \\
        y_{2} \\
        \vdots \\
        y_{L}
    \end{array} \right)  ,
\end{equation}
so Eq. \eqref{y-linear-reg} can be rewritten as $\mathbf{y} = V \mathbf{W}$. If we want the output $\mathbf{y}$ to get as close as possible to the desired target function, $\bar{\mathbf{y}}$, we choose the set of weights in order to minimize the mean square error (MSE),
\begin{equation} \label{mse-eq}
    \text{MSE}\left( \mathbf{y}, \bar{\mathbf{y}} \right) = \frac{1}{L} \sum_{k=1}^{L} \left( y_{k} - \bar{y}_{k} \right)^{2}  .
\end{equation}
It can be shown that the optimal set of weights to reach the minimum of Eq. \eqref{mse-eq} are the ones obtained through $\mathbf{W}_{\text{opt}} = V^{\text{MP}} \bar{\mathbf{y}}$,
where $V^{MP} = \left(V^{\top} V\right)^{-1} V^{\top}$ is the Moore-Penrose inverse of $V$ \cite{LUKOSEVICIUS2009127}. The higher the value of $L$ the more precise our estimation of the optimal weights will usually be. Once the system has been trained, we consider a (smaller) test set of $L^{\prime}$ inputs to be fed into the reservoir afterwards. The RC performance is then checked with this test set of new unseen data using the MSE metric from Eq. \eqref{mse-eq}.

\subsection{Description of the platform} \label{platform-sect}
\begin{figure}[h!]
    \centering
    \includegraphics[width=0.8\linewidth]{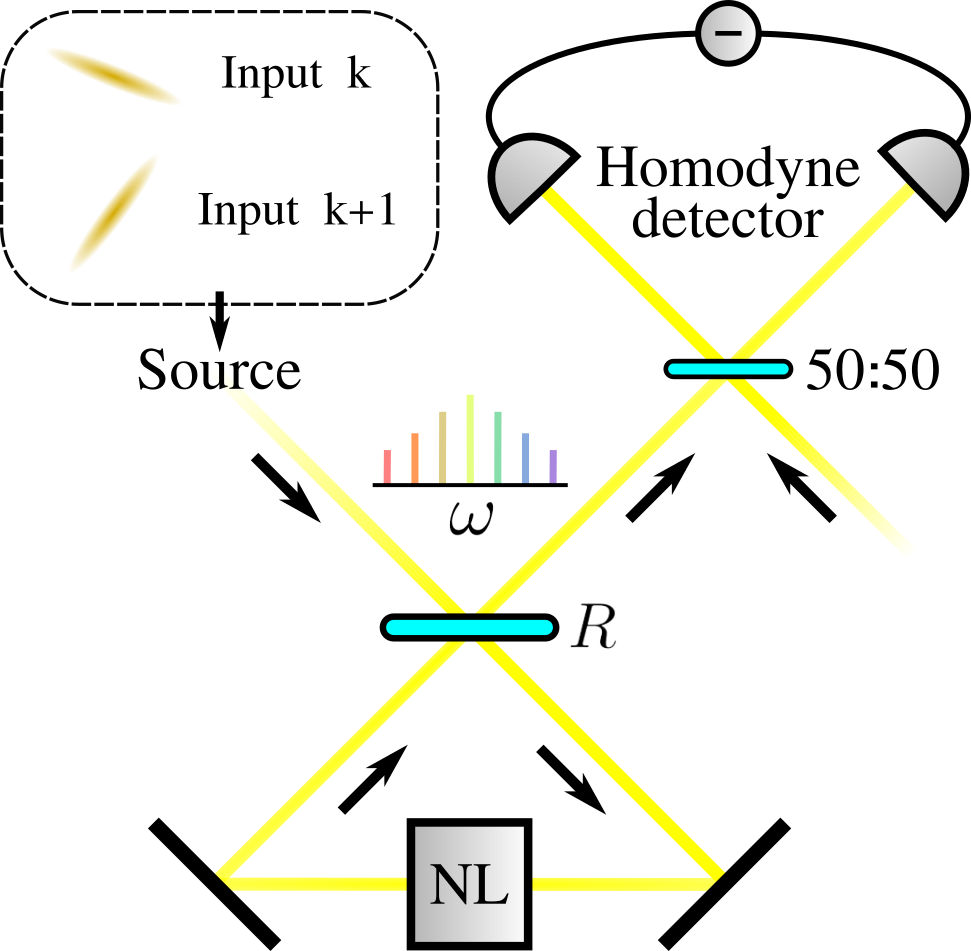}
    \caption{Scheme of the loop-based architecture.}
    \label{figure-1}
\end{figure}

Our architecture works in the CV quantum optical regime. The physical substrate is an $N$-mode optical pulse traveling through a closed optical loop or cavity ($N$ denotes the size of the reservoir). The $N$-mode internal degrees of freedom inside each pulse can be attained via, for example, frequency multiplexing \cite{Araujo2014,Roslund2014,60-mode-frequency,Cai2017b,kouadou2022spectrally}. In optics, frequency multiplexing has already been shown to be a useful strategy for classical RC \cite{butschek2022photonic}. In our approach, the external information is injected from a pulse-generating light source, which provides squeezed vacuum states. The input sequence is encoded in the squeezing phases of the input pulses (one input value for each pulse), depicted as \textit{source} in Fig. \ref{figure-1}. Fast, accurate, and reconfigurable phase-setting devices have already been used in experiments with great impact \cite{Madsen2022}. Each input pulse is coupled to the loop pulse using a beam-splitter (BS), with reflectivity $R$ shown in Fig. \ref{figure-1}, yielding two output pulses. One of them remains in the loop and gives feedback to the next iteration (creating a quantum memory). In this way, the reservoir can retain information from previous inputs without the need for external memory. The remaining output pulse is passed to a detector that measures each mode and uses the obtained observables for the readout layer. The fraction of light that remains in the cavity on each round trip is determined by the BS reflectivity $R$.

Inside the cavity, a nonlinear medium is placed (\textit{NL} in Fig. \ref{figure-1}), which applies a dynamical transformation to the loop pulse each time it passes through. This creates a complex optical network \cite{Roslund2014,Nokkala_2018,Cabot2018,Cai2017b,renault2023experimental} and can be modeled by a Hamiltonian that is quadratic in the field operators,
\begin{equation} \label{general-hamil}
    \hat{H} = \frac{1}{2}\sum_{i,j = 1}^{N} \left( \alpha_{ij} \hat{a}_{i}^{\dagger} \hat{a}_{j} + \beta_{ij} \hat{a}_{i}^{\dagger} \hat{a}_{j}^{\dagger} + \text{h.c.} \right) ,
\end{equation}
where $\hat{a}_{i}$ ($\hat{a}_{i}^{\dagger}$) is the annihilation (creation) operator of mode $i$. The coupling terms $\alpha_{ij}$ and $\beta_{ij}$ encode different network topologies and lead to entanglement among modes inside the loop pulse, at each round trip. If all the terms $\beta_{ij} = 0$, the dynamical transformation is called \textit{passive}, whereas if there are any $\beta_{ij} \neq 0$ the transformation is \textit{active}. Active transformations in CV quantum optics do not conserve the average number of photons of the quantum state and are known to generate squeezing, the main resource for entanglement \cite{adesso2014, serafini2017}. It is important to note that a passive cavity also produces entangled states because the external input pulses are already squeezed (even though it does not generate additional squeezing).

The detector from Fig. \ref{figure-1} performs homodyne measurements to every mode in the incoming pulses. Concretely, the measured operator are the $x$-quadrature $\hat{x}_{i} = \frac{1}{\sqrt{2}} \left( \hat{a}_{i} + \hat{a}_{i}^{\dagger} \right)$ ($i = 1, 2, \dots, N$). From these measurements, the moments of the operator vector $\hat{\mathbf{x}} = \left( \hat{x}_{1}, \dots, \hat{x}_{N} \right)^{\top}$ can be computed and used as observables for the readout layer. As we are injecting squeezed vacuum states, only even-order moments are considered. For the tasks shown in Sect. \ref{sec-III}, the chosen set of observables is composed of second and fourth-order moments. Concretely, the chosen set is $\left\{ \left\langle \hat{x}_{i} \hat{x}_{j} \right\rangle, \left\langle \hat{x}_{i}^{2} \hat{x}_{j}^{2} \right\rangle, \left\langle \hat{x}_{i}^{3} \hat{x}_{j} \right\rangle \right\}_{i,j = 1}^{N} \equiv \left\{ O_{l} \right\}_{l=1}^{N (3N+1)/2}$, which has a total of $N (3N + 1)/2$ observables. In Gaussian states, fourth-order moments can be written as nonlinear functions of the second-order moments. In our case, they are useful for enhancing the accessible nonlinear terms, which are relevant for several tasks and come at no experimental expenditure \cite{garciabeni2023}. We note that the readout size scales quadratically with the number of modes $N$. Usually, in optical reservoir computing the dimensionality scales linearly with $N$, as the information is encoded in field amplitudes \cite{larger2017high,VanDerSande2017,mackey-glass-forecasting}. By introducing the inputs in the field quantum fluctuations we access a broader dimensional space by generating mode correlations and entanglement, which allows more complex information processing in relatively smaller reservoirs \cite{Nokkala2021,garciabeni2023}. To access the averaged values of the field correlations we consider averaging over an ensemble of realizations \cite{Fujii-Nakajima}.

Each RC time step encompasses the whole process we have detailed: BS coupling of the input and loop pulses, detection of the output pulse, and transformation of the loop pulse under the non-linear crystal. There are as many time steps as samples in the input sequence, and they will be labeled with the letter $k$. So at the $k$-th time step, the input $s_{k}$ will be encoded and introduced into the system and the vector of observables $\mathbf{O}^{(k)} = \left( O_{1}^{(k)}, O_{2}^{(k)}, \dots \right)$ is measured and used for the readout layer.

The platform is based on the proposal in Ref. \cite{garciabeni2023}, where real-time information processing was reported. The main novelty here is that we specifically address the role of a squeezing reservoir, with both active and passive transformation. The goal is to assess the importance of quantum resources for the performance of QRC.
In order to simplify the experimental footprint consider a single NL crystal. Indeed this has not significant effect when processing past inputs.
This design can be also adapted for single-loop ensemble processing adding a fiber, as shown in \cite{garciabeni2023}.

\section{Reservoir computing tasks under additive noise} \label{sec-III}
Noise in the readout layer is known to be significantly detrimental to RC performance \cite{Soriano_noise_1,Soriano_noise_2}. Some strategies have been developed in different architectures to make reservoirs more robust to noise \cite{Soriano_noise_1,Johannes_fluctuations}. In this section, we study the effect of the amount of squeezing produced by the cavity crystal on the noise robustness of the platform in the readout layer and compare it to the one obtained by tuning the BS reflectivity. In our simulations, readout noise is included as additive fluctuations in the measured observables, $\mathbf{O}_{\text{meas}}^{(k)}$, concretely
\begin{equation}
\mathbf{O}_{\text{meas}}^{(k)} = \mathbf{O}_{\text{ideal}}^{(k)} + \mathbf{\mathcal{E}}^{(k)}(0, \sigma_{\text{noise}}^{2})  ,
\end{equation}
where $\mathbf{O}_{\text{ideal}}^{(k)}$ stands for the observable that would be measured in the ideal case of zero fluctuations, which would correspond to an infinite number of measurements, and $\mathbf{\mathcal{E}}^{(k)}$ stands for the additive noise vector. For the noise, we model it as normally distributed fluctuations of variance equal to $\sigma_{\text{noise}}^{2}$ applied to the measured quadratures. Although the added noise is absolute (it does not depend on the magnitude of the observables),  we can use the vacuum noise variance ($\sigma_{\text{vacuum}}^{2} = 1/2$ in our case) or shot noise as a relative measure of the additive noise intensity. In that regard, added noise of variance equal to $0.1$ would be equivalent to $20\%$ of vacuum fluctuations.

In our simulations, we generate every crystal Hamiltonian, Eq. \eqref{general-hamil}, randomly with the condition that every one of its supermodes is squeezed by
$e^{-r}$ (see App. \ref{App-1} for details). In every realization, the modes of the input pulses are squeezed with a fixed squeezing strength, $r_{\text{input}} = 2$ (approximately 8.7 dB). The encoding function of the squeezing phase, $\phi_{k} = f(s_{k})$, is tuned depending on the task we are considering. Concretely, we consider the family of linear functions $\phi_{k}^{(m)} = m \pi s_{k}$. In this respect, the smaller the value of $m$, the better the reservoir is at reproducing linear and quadratic functions of $s_{k}$. For increasing values of $m$, higher non-linear contributions become more relevant \cite{garciabeni2023,Nokkala2021}.

We consider three temporal tasks: the linear memory task, the \textit{nonlinear autoregressive moving average} (NARMA) task, and the forecasting of the Mackey-Glass chaotic time series \cite{mackey-glass-chaotic-1, mackey-glass-chaotic-2}. These three tasks provide a broad overall picture of the properties of the reservoir memory for both linear and nonlinear computations. After applying the training protocol described in Sect \ref{RC-section} for a given target function, we check the performance of the trained reservoirs on an additional test input sequence. For evaluation, we mainly use the mean-square error, Eq. \eqref{mse-eq}, after training optimization, $\text{min}_{\mathbf{W}} \text{MSE}\left( \mathbf{y}, \bar{\mathbf{y}} \right)$. To get the MSE to be normalized between 0 and 1, we normalize both the target and the reproduced function data to zero mean and one standard deviation.

\subsection{Linear memory task} \label{sec-IIIA}
The linear memory task is the simplest way to check the accessible memory of our reservoir in the presence of noise. We train the reservoir to reproduce past entries from the input series. That is, we consider the target function
\begin{equation} \label{linear-target-function}
    \bar{y}_{k}(\mathbf{s},d) = s_{k-d}  ,
\end{equation}
where we aim to to reproduce, at time step $k$, the input that was introduced $d$ time steps in the past (or the input at \textit{delay} $d$). For simulations, we consider an input sequence composed of random entries from a uniform distribution in the interval $[-1, 1]$. We tune the squeezing angle encoding to be $\phi_{k} = \pi s_{k}/4$ to maximize linear contributions. For visualization purposes, we use the linear capacity, defined as $C(d) = 1 - \text{min}_{\mathbf{W}} \text{MSE}\left( \mathbf{y}, \bar{\mathbf{y}}(d) \right)$, to test the performance of this task. The vector $\bar{\mathbf{y}}(d)$ is composed of the target function from Eq. \eqref{linear-target-function} at each time step as a function of the delay $d$. Training and test set sizes are set to be 4000 and 1000, respectively.
\begin{figure}[h!]
    \centering
    \includegraphics[width=\linewidth]{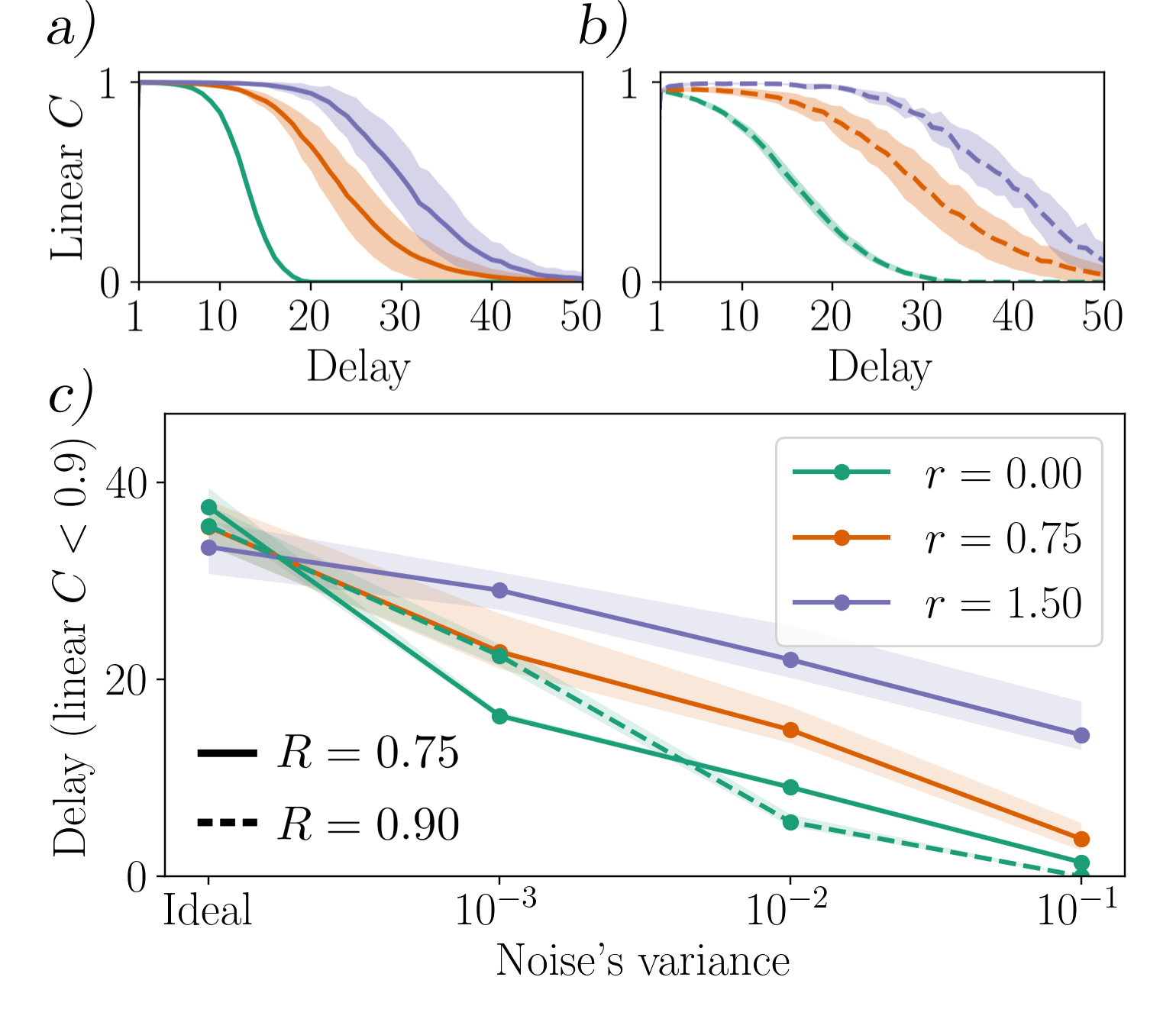}
    \caption{\textbf{Linear memory under additive noise:} (a-b) linear capacity as a function of the delay for different values of squeezing (green for $r = 0$, orange for $r = 0.75$ and purple for $r = 1.5$) and different reflectivities: $R = 0.75$ in (a) and $R = 0.9$ in (b). In both figures, the noise variance is $\sigma_{\text{noise}}^{2} = 10^{-2}$. The curves are taken from averaging among 100 different random realizations and the shadows depict the standard deviation. (c) delay at which the linear capacity drops below $0.9$ as a function of $\sigma_{\text{noise}}^{2}$, including the ideal case ($\sigma_{\text{noise}}^{2} = 0$) for different values of squeezing (different colors) and different values of the reflectivity (line format).}
    \label{figure-2}
\end{figure}

Figures \ref{figure-2}a and \ref{figure-2}b show the linear capacity as a function of the delay $d$ on the target function from Eq. \eqref{linear-target-function} for different values of the reflectivity ($R=0.75$ in Fig. \ref{figure-2}a and $R=0.9$ in Fig. \ref{figure-2}b). In both plots, the noise variance is $10^{-2}$ ($2\%$ of vacuum fluctuations). We see that for both reflectivities, increasing the cavity squeezing improves the attainable memory. Having a higher reflectivity provides a longer `tail' in the linear capacity at the expense of reducing it for small and intermediate delays. This is the effect of the smaller amount of light leaving the cavity for increasing values of $R$. In Fig. \ref{figure-2}c the delay at which the linear capacity drops below $0.9$ (which we also call the \textit{delay cut}) is plotted as a function of the noise variance. We find that cavity squeezing provides significant noise robustness. For the reflectivity $R = 0.75$, adding cavity squeezing equal to $r = 1.5$ provides a high linear capacity beyond delay $10$ for a noise intensity of $\sigma_{\text{noise}}^{2} = 0.1$. In Fig. \ref{figure-2}c the drawback of increasing the BS reflectivity can also be noted: for small noise intensity ($\sigma_{\text{noise}}^{2} = 10^{-3}$) and no cavity squeezing (light color), having $R = 0.9$ (dashed line) improves the delay cut compared to $R = 0.75$ (solid line), as increasing the reflectivity improves the memory robustness to noise. However, when the noise intensity is increased, the delay cut for $R = 0.9$ drops below the one for $R = 0.75$. This is because the higher the reflectivity, the less light leaves the cavity and travels to the detector. If the readout fluctuations are large compared to the intensity of the light coming from the loop, the accessible memory of the reservoir will be severely degraded. 
The cavity squeezing increases significantly the accessible linear memory in the presence of a large noise intensity.

\subsection{NARMA10 task} \label{sec-IIIB}
In this section, we are analyzing the performance of the reservoir for the NARMA task, which requires high linear and non-linear memory. It is one of the most common benchmark tasks and has been used to test several QRC proposals \cite{rodrigo_PRL,Johannes_fluctuations}. In this article, we consider the NARMA10 task, with a target function at time $k$ being
\begin{eqnarray} \label{NARMA-target-eq}
    \bar{y}_{k} &=& \alpha \bar{y}_{k-1} + \beta \bar{y}_{k-1} \sum_{i=1}^{10} \bar{y}_{k-i} + \gamma u_{k-1} u_{k-10} + \delta  , \\
    u_{k} &=& \mu + \nu s_{k}  ,
\end{eqnarray}
where the default constant parameters are set to $(\alpha, \beta, \gamma, \delta) = (0.3, 0.05, 1.5, 0.1)$. The function input parameters $(\mu,\nu)$ are chosen to be $(0,0.2)$, where $s_{k}$ are taken from a uniform random distribution in the interval $[-1, 1]$ (they are the source inputs). For the reservoir to be able to have a good performance on the NARMA10 task it is necessary for it to be able to have a high linear capacity up to delay 10 and a low error reproducing the function $y_{k} = s_{k-1} \cdot s_{k-10}$ \cite{narma10-anatomy}.

\begin{figure}[h!]
    \centering
    \includegraphics[width=\linewidth]{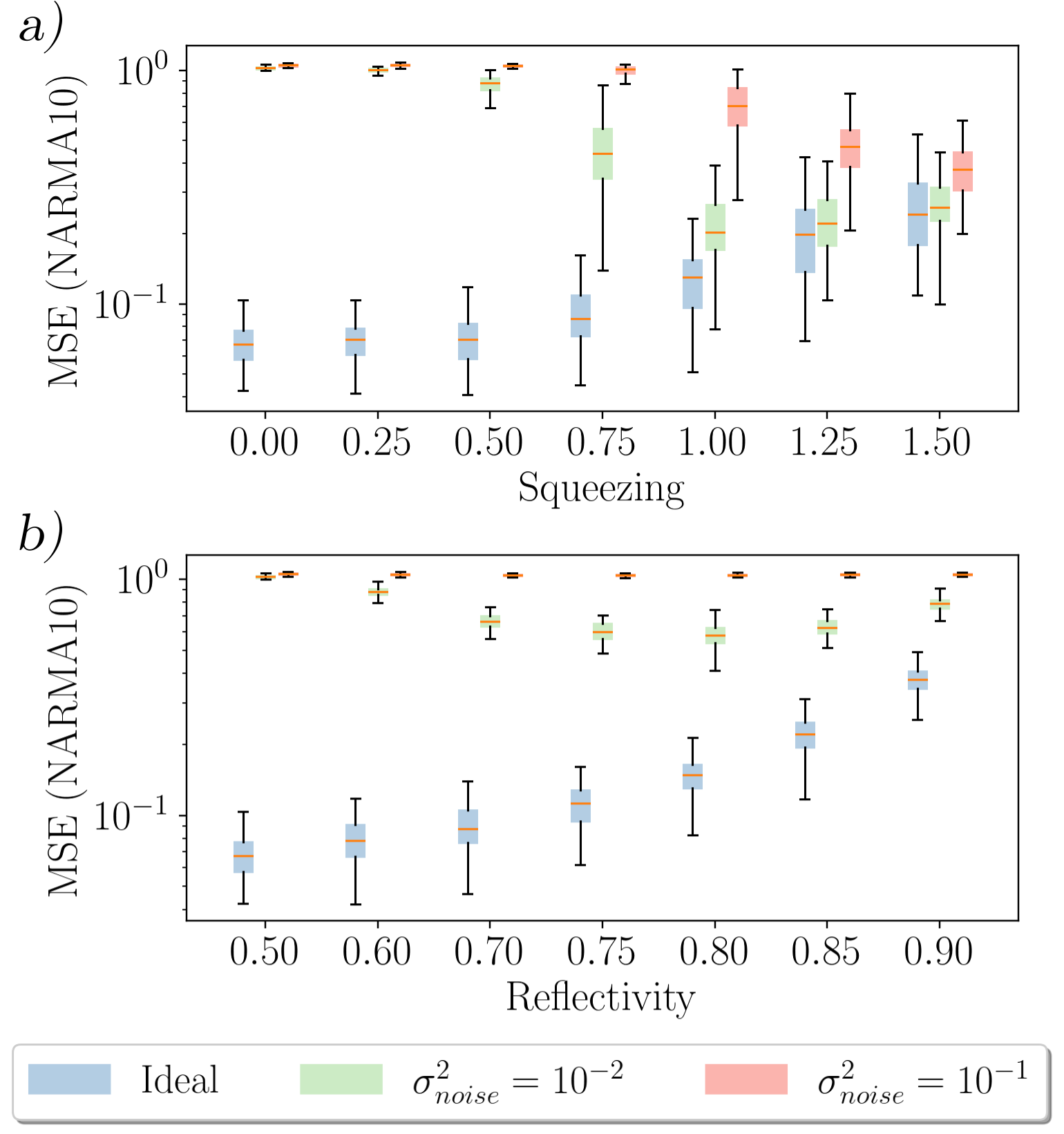}
    \caption{\textbf{Performance on NARMA10 task:} box plot of the mean square error (MSE) of the NARMA10 task as a function of (a) squeezing and (b) reflectivity for different values of the noise variance (different colors). For Fig. \ref{figure-3}a we take the reflectivity equal to $R = 0.5$ and in Fig. \ref{figure-3}b the cavity squeezing is equal to zero. For a given value of the x-axis, the boxes for each noise scenario are split to avoid overlapping.}
    \label{figure-3}
\end{figure}

To perform this task we consider the same input encoding as in Sect. \ref{sec-IIIA}, namely $\phi_{k} = \pi s_{k}/4$. To test the performance we use the MSE defined in Eq. \eqref{mse-eq}. In Figs. \ref{figure-3}a and \ref{figure-3}b the performance is shown, comparing the effect of cavity squeezing (x-axis in Fig. \ref{figure-3}a) and BS reflectivity (x-axis in Fig. \ref{figure-3}b). Three different noise scenarios are considered in both figures: the ideal case (blue boxes) and noise variance equal to $10^{-2}$ (green boxes) and $10^{-1}$ (pink boxes). In the ideal case, the optimal values of cavity squeezing and BS reflectivity are found to be $r = 0$ and $R = 0.5$. It can be seen that increasing either cavity squeezing or reflectivity degrades the performance in the absence of noise. The reason for this is that, for the chosen encoding and observables, increasing these two parameters increases linear memory at the expense of quadratic memory, which is also very relevant for the NARMA task. This balance among linear and non-linear memory is well known in the field of RC \cite{dambre2012information,inubushi2017reservoir,Nokkala2021}.
\\
\\
Moving to the realistic case of a finite number of measurements, for a noise intensity of $\sigma_{\text{noise}}^{2} = 10^{-2}$, the improvement of the cavity squeezing over the reflectivity is clearly seen. In that noise scenario, increasing both parameters improves the performance up to an optimal value ($r \sim 1$ and $R \sim 0.8$). However, in the passive cavity case, the optimal value has a higher error in comparison to the active cavity and is further away from the ideal case error. For higher values of the noise, $\sigma_{\text{noise}}^{2} = 10^{-1}$, most examples completely fail at attempting to reproduce the NARMA function. Only when the cavity squeezing $r \gtrsim 1$, the MSE drops below 1. This is due to the fact that in most scenarios the noise level makes it impossible for the reservoir to resolve inputs with a delay of 10 or more as its intensity becomes comparable with the value of the system observables, giving a bad signal-to-noise ratio \cite{garciabeni2023}.

Even though there is a counterbalance between linear and quadratic memory, and thus increasing the cavity squeezing and the BS reflectivity is detrimental to the performance in ideal scenarios, in the presence of readout noise the active cavities with high squeezing outperform the rest. Indeed, the only case where the performance of the noisy reservoir comes close to the ideal case is when the cavity squeezing is higher than $r = 1$. This provides an interesting example where the role of a given resource needs to be addressed beyond ideal settings, as benefits could arise in the presence of noise.
\subsection{Time series prediction of a chaotic signal} \label{sec-IIIC}
One of the main applications of RC is time series forecasting, and thus in this section, we consider the task of forecasting the Mackey-Glass time series \cite{mackey-glass-original}. The differential equation that describes the dynamics of the signal is
\begin{equation} \label{mackey-glass-diffeq}
    \dot{s}(t) = -0.1 s(t) + \frac{0.2 s(t - \tau)}{1+ s(t - \tau)^{10}}  ,
\end{equation}
where for $\tau = 17$ the time series is chaotic \cite{mackey-glass-chaotic-1,mackey-glass-chaotic-2}. For the input sequence, we have sampled the solutions to Eq. \eqref{mackey-glass-diffeq} with time resolution $t_{r} = 3$ \cite{mackey-glass-forecasting}, so that $s_{k} = s(t_{0} + k t_{r})$ (the initial conditions are chosen randomly). The target function for the training will be to predict the next input in the sequence, that is, $\bar{y}_{k} = s_{k+1}$. For this specific task, we use the input encoding $\phi_{k} = \pi s_{k}$, which provides higher nonlinear memory \cite{garciabeni2023}. Once the reservoir has been trained to predict the next value of the signal, we can feed the predicted values as new inputs for the protocol. The reservoir thus, ideally, faithfully reproduces the chaotic signal without the need for new input data. We call this protocol \textit{autonomous driving}.

\begin{figure}[h!]
    \centering
    \includegraphics[width=\linewidth]{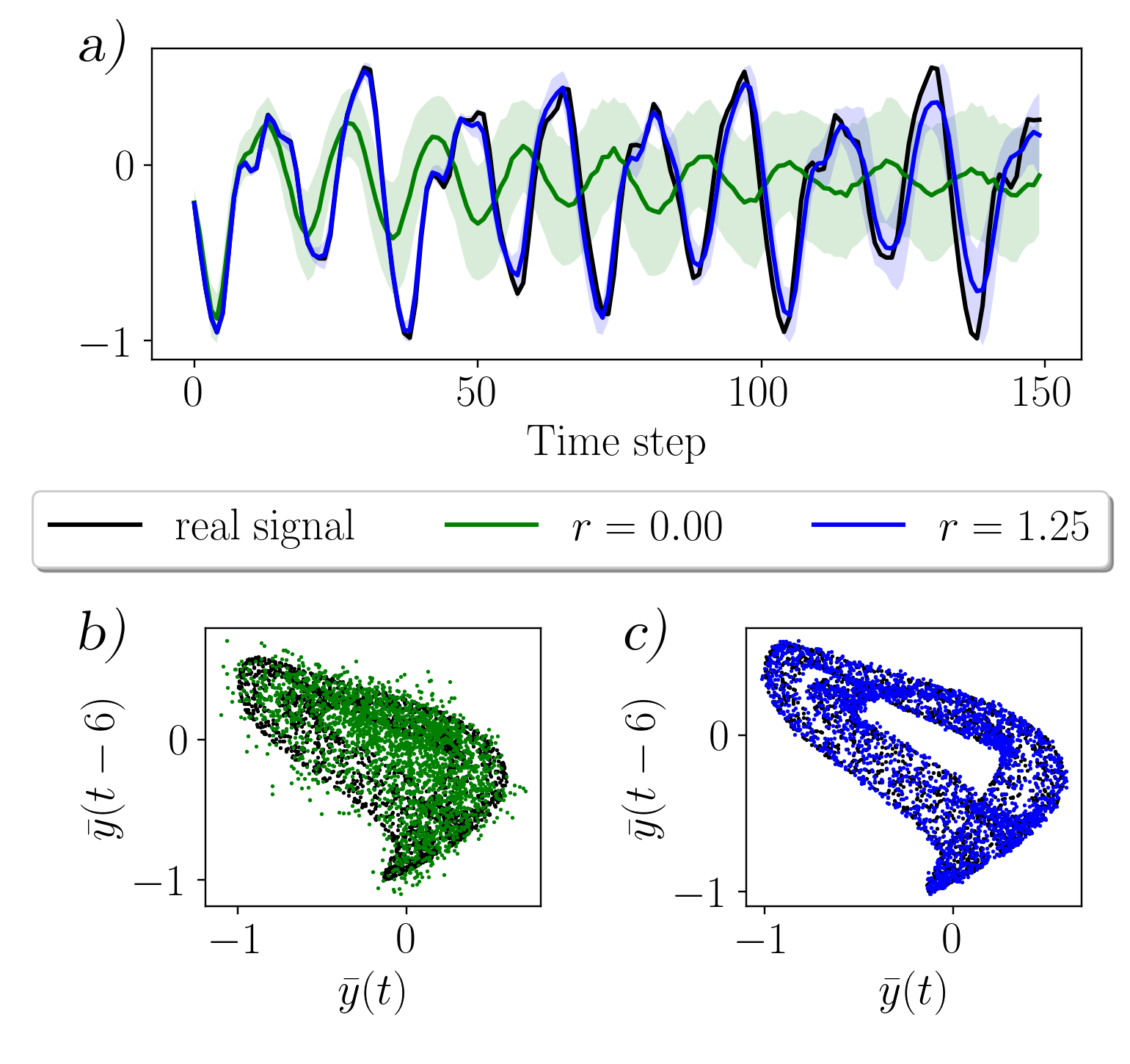}
    \caption{\textbf{Mackey-Glass time series prediction under additive noise:} (a) time series as a function of the reservoir time steps: the black curve shows the real-time series while the green and blue curves show the autonomously driven predictions for $r = 0$ and $r = 1.25$, respectively (averages taken among 100 realizations, shadows depicting the standard deviation). (b-c) chaotic attractor in the phase space $y(t)$-$y(t−6)$: the black points provide the real attractor while the green and blue dots provide the results of an autonomously driven realization with $r = 0$ and $r = 1.25$, respectively. In every simulation, the BS reflectivity is set to $R = 0.75$ and the noise variance is $\sigma_{\text{noise}}^{2} = 10^{-1}$.}
    \label{figure-4}
\end{figure}

In Fig. \ref{figure-4}a the autonomously driven signal evolution is plotted for two different values of the cavity squeezing (green curve for $r = 0$ and blue curve for $r = 1.25$) while the noise variance is kept at $\sigma_{\text{noise}}^{2} = 10^{-1}$. The real signal is shown as a black curve for comparison. We see that the active cavity performs a much better prediction than the passive one in the long term, achieving accurate predictions up to 50 time steps. The passive cavity reservoir cannot overcome the effects of noise, and thus the prediction performance drops dramatically. In Figs. \ref{figure-4}b and \ref{figure-4}c, we compare the Mackey-Glass chaotic attractor, black dots, to $3000$ values from single realizations of trained reservoirs with a passive cavity (Fig. \ref{figure-4}b, green dots) and an active cavity with $r = 1.25$ (blue dots). Also in this case, we see that the reservoir with cavity squeezing is able to approximately reproduce the attractor, while the passive reservoir is not.
\section{Accessible memory enhancement} \label{sec-IV}
In this section, we will explain in detail the reason behind the performance improvement due to the cavity squeezing that we have shown in the previous sections. From a physical point of view, the BS and the cavity crystal can induce competing effects on the reservoir memory. The BS causes a loss of photons in the loop pulse. On the other hand, the non-linear processes that arise from the interactions inside the crystal can increase the energy of the loop pulse. Concretely, active crystals (those that produce squeezing) increase the total photon number inside the pulse, as opposed to passive crystals, which maintain it constant. This energy enhancement from active crystals counteracts BS losses and helps retain information inside the loop pulse for longer times.

\begin{figure}[h!]
    \centering
    \includegraphics[width=\linewidth]{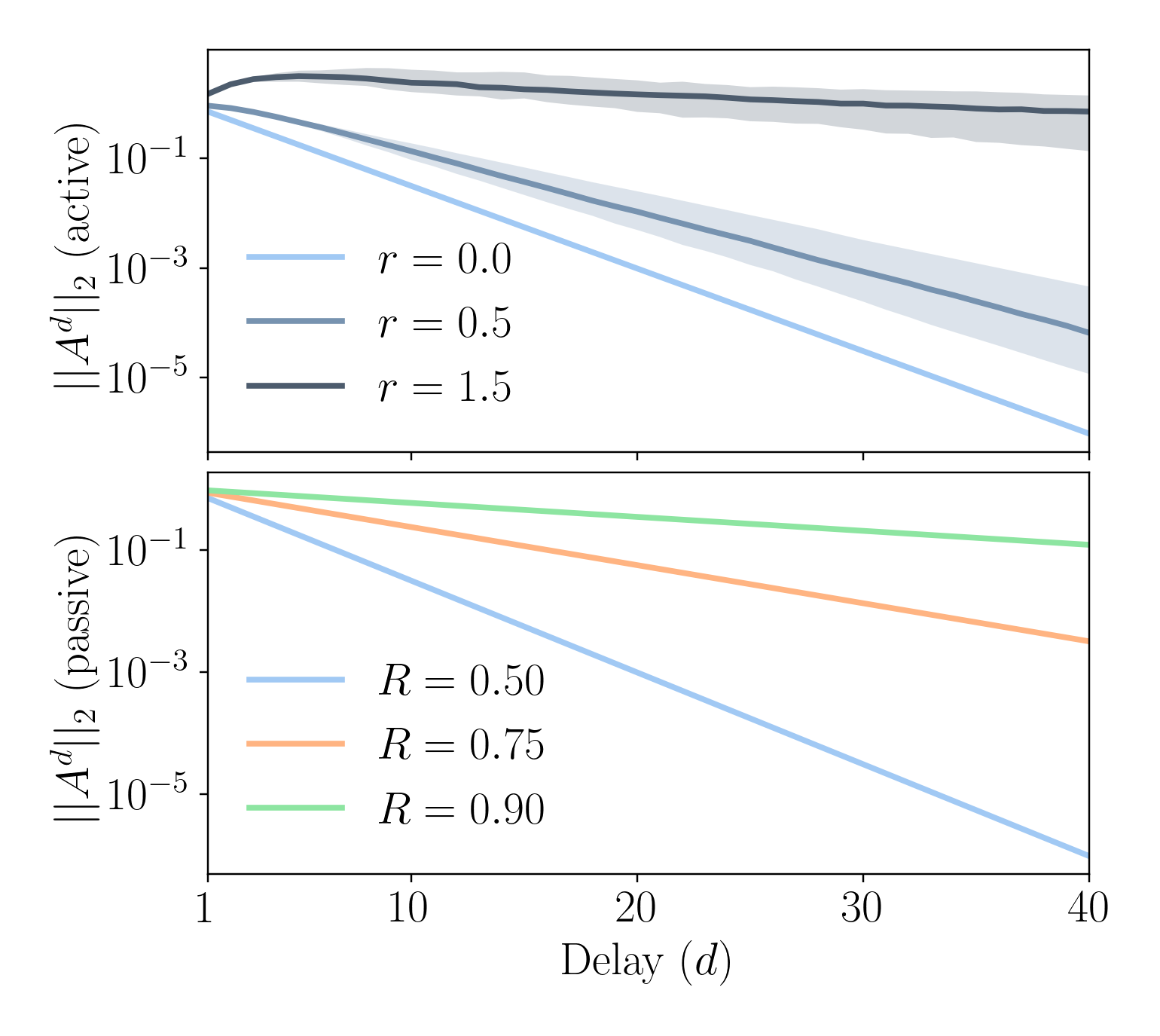}
    \caption{\textbf{Spectral norm of loop matrix dynamics:} spectral norm of the loop dynamical matrix $A$ to the power of the delay $d$ for (a) active transformations with $R = 0.5$ and different values of the squeezing, $r$, and (b) passive transformations ($r = 0$) for different values of the reflectivity. Each curve shows the median from 100 different realizations while the shades range from the first to the ninth decile.}
    \label{figure-5}
\end{figure}

To quantify how these effects contribute constructively to the functioning of the QRC, we study the time evolution of the loop pulse during the protocol. At each round trip, the field quadratures of the loop pulse transform via the symplectic matrix $A = \sqrt{R} S$, where $R$ stands for the BS reflectivity and $S$ is the symplectic matrix modeling the evolution of the pulse inside the cavity crystal (see Apps. \ref{App-1} and \ref{App-2} for details on the matrices $S$ and $A$, respectively). The memory retention of our reservoir is directly related to the powers of the symplectic matrix $A$ (App. \ref{App-recursive-eqs}) and can be quantified using the spectral norm of $A^{d}$ (written as $\lVert A^{d} \rVert_{2}$), in which $d$ resembles the delay of the input information (Eq. \eqref{spectral-norm} from App. \ref{App-memory-decay}). In that regard, the faster $\lVert A^{d} \rVert_{2}$ decays to zero, the smaller the memory retention of our reservoir will be (and vice-versa). It can be analytically shown that, if the crystal inside the cavity is passive, then $\lVert A^{d} \rVert_{2} = R^{d/2}$ and, if it is active, then $\lVert A^{d} \rVert_{2} \geq R^{d/2}$. From these equations, we infer that active cavity crystals can improve the memory retention of the reservoir (similarly to the direct effect of the BS reflectivity). In Fig. \ref{figure-5} we show the values of $\lVert A^{d} \rVert_{2}$ as a function of the delay for the randomly generated Hamiltonians that we considered in Sect. \ref{sec-III}. In Fig. \ref{figure-5}a the spectral norm of $A$ is plotted for active cavities with different values of cavity squeezing, $r$, while in Fig. \ref{figure-5}b the same is shown for passive cavities with different values of the BS reflectivity. We see how the negative slope of the spectral norm decreases as we increase either the cavity squeezing (Fig. \ref{figure-5}a) or the BS reflectivity (Fig. \ref{figure-5}b). This means that the magnitude of the delayed input information decays more slowly, and thus it may be reproduced by our trained reservoir more easily. In the presence of readout noise, input information decaying more slowly with time translates into more accessible memory and improves the reservoir's robustness to noise This is the reason why cavity squeezing improves the performance in every benchmark task shown in Sect. \ref{sec-III}. While both $R$ and $A$ contribute to such a memory enhancement, increasing the BS reflectivity has the drawback of reducing the cavity light reaching the detector (lowering the signal-to-noise ratio) and thus is not as effective for increasing the noise robustness.
\section{Conclusion} \label{conclusions}
In light of the significant achievements in classical RC \cite{brunner2019photonic}, photonic platforms are emerging as promising candidates for quantum implementations. Commendable features are, for instance, fast processing rates and low decoherence at room temperature \cite{Mujal_opportunities}. Different photonic QRC platforms have already been theoretically explored and show improvements due to the enlarged Hilbert space \cite{Nokkala2021, Spagnolo2022} as well as the ability of real-time processing without the use of external memories \cite{garciabeni2023}.

Detrimental effects of readout noise have been discussed both in classical RC \cite{Soriano_noise_1,Soriano_noise_2} and in QRC settings \cite{Johannes_fluctuations,time-series-QRC-measurements,garciabeni2023}. In the case of quantum reservoirs, the problem is even more profound, as readout noise is theoretically unavoidable due to the stochastic nature of quantum measurements, which produce statistical fluctuations that can hinder any possible quantum advantage \cite{Mujal_opportunities, time-series-QRC-measurements, garciabeni2023}. In this paper, we demonstrate the performance-enhancing potential of quantum squeezing (active cavity) applied to a vacuum state quantum memory (loop pulse) to overcome noise in realistic scenarios.
This establishes squeezing as a quantum resource for accessing the enlarged space of quantum correlations and entanglement and for improving the performance in relevant benchmark tasks, either predictive or requiring memory.
Even though tuning the BS reflectivity also improves memory retention, as in \cite{garciabeni2023}, increasing cavity squeezing is shown to be a preferred method to improve the reservoir robustness under adverse noise conditions.
Interestingly, the effect of squeezing on the QRC performance when accounting for measurement noise completely deviates from predictions under ideal conditions.

In summary, state-of-the-art frequency multiplexed quantum networks \cite{60-mode-frequency,Cai2017b,kouadou2022spectrally,renault2023experimental} 
represent a powerful set-up for near-term experimental implementations of QRC. Our results serve as a  guide for the experimental design laying the foundations for photonic QRC in CV in realistic noisy scenarios and exploiting quantum resources.

\begin{acknowledgments}
We acknowledge the Spanish State Research Agency through the  QUARESC project (PID2019-109094GB-C21 and -C22/ AEI / 10.13039/501100011033) and through the Severo Ochoa and Mar\'ia de Maeztu Program for Centers and Units of Excellence in R\&D (CEX2021-001164-M funded by the MCIN/AEI/10.13039/501100011033) and. We also acknowledge funding by CAIB through the QUAREC project (PRD2018/47). J.G-B. is funded by the Conselleria d’Educació, Universitat i Recerca of the Government of the Balearic Islands with grant code FPI/036/2020. G.L.G. is funded by the Spanish  MEFP/MiU and co-funded by the University of the Balearic Islands through the Beatriz Galindo program  (BG20/00085). This work has been financially supported by MINECO through the QUANTUM
ENIA project, and by EU through the RTRP - NextGenerationEU within the framework of the Digital Spain 2026 Agenda. The CSIC Interdisciplinary Thematic Platform (PTI) on Quantum Technologies in Spain is also acknowledged.
\end{acknowledgments}
\appendix
\section{Symplectic formalism of the crystal dynamics} \label{App-1}
A CV quantum state is completely determined, in the Heisenberg picture, by the statistics of the quadrature vector $\hat{\mathbf{Q}} = \left( \hat{x}_{1}, \hat{p}_{1}, \dots, \hat{x}_{N}, \hat{p}_{N} \right)^{\top}$, where $\hat{x}_{i} = \frac{1}{\sqrt{2}} \left( \hat{a}_{i}^{\dagger} + \hat{a}_{i} \right)$ and $\hat{p}_{i} = \frac{i}{\sqrt{2}} \left( \hat{a}_{i}^{\dagger} - \hat{a}_{i} \right)$ ($i = 1, \dots, N$) are, respectively, the amplitude and phase quadratures of each mode. For quadratic Hamiltonians as in Eq. \eqref{general-hamil} the evolution of the quadrature vector can be written as
\begin{equation}
    e^{i \hat{H} t} \hat{\mathbf{Q}} e^{-i \hat{H} t} = S_{\hat{H}}(t) \hat{\mathbf{Q}}  ,
\end{equation}
where $S_{\hat{H}}$ is a $2N \times 2N$ symplectic matrix \cite{adesso2014, serafini2017}. For the sake of clarity, we will drop the $\hat{H}$ subscript (and the time dependency) and call the symplectic matrix simply as $S$. Every symplectic matrix admits a Bloch-Messiah decomposition \cite{bloch-messiah,bloch-messiah-reexamination}, 
\begin{eqnarray} \label{bloch-messiah-S}
    S &=& U \Delta_{\mathbf{\xi}} V  , \\ \label{squeezing-supermodes}
    \Delta_{\mathbf{\xi}} &=& \text{diag}\left( \xi_{1},\xi_{1}^{-1}, \xi_{2},\xi_{2}^{-1}, \dots, \xi_{N},\xi_{N}^{-1} \right)  ,
\end{eqnarray}
where $U$ and $V$ are orthogonal symplectic matrices and $\xi_{i}^{-1}$ ($\xi_{i}$) provides the squeezing (anti-squeezing) applied to the $i$-th supermode (we consider $\xi_{i} \geq 1 \ \forall i$). Thus the squeezing and anti-squeezing values obtained from the Bloch-Messiah decomposition correspond to the singular values of $S$. So the transformation is passive iff all the singular values of $S$ are equal to 1 ($\Delta_{\mathbf{\xi}}$ is equal to the identity). It is trivial to show that if $S$ is passive, then all its powers $S^{d}$ are also passive.

For the simulations we have performed in this manuscript, we have considered the non-linear crystal squeezes every supermode by the same amount, so $\xi_{i}^{-1} \equiv e^{-r/2}$ ($\forall i$), where $r$ denotes the squeezing strength per mode applied by the Hamiltonian. So for our simulations we set $\Delta_{\xi} = \bigoplus_{i=1}^{N} \text{diag} \left( e^{r/2}, e^{-r/2} \right)$ and choose the matrices $U$ and $V$ randomly to generate the symplectic matrix $S$, applying Eq. \eqref{bloch-messiah-S}.
\section{QRC dynamics} \label{App-2}
\subsection{Recursive equations} \label{App-recursive-eqs}
In this section, we will work out the equations describing the reservoir dynamics to gain insight into the effect of cavity squeezing. We start by considering the quadrature operators of the pulses after the BS coupling at time step $k$. We write them as
\begin{eqnarray} \label{out-R-eq}
    \hat{\mathbf{Q}}_{\text{out}}^{(k)} &=& \sqrt{1-R} \hat{\mathbf{Q}}_{\text{loop}}^{(k)} - \sqrt{R} \hat{\mathbf{Q}}_{\text{in}}^{(k)}  , \\ \label{loop-R-eq}
    \hat{\mathbf{Q}}_{\text{loop}}^{(k) \prime} &=& S \left(\sqrt{R} \hat{\mathbf{Q}}_{\text{loop}}^{(k)} + \sqrt{1-R} \hat{\mathbf{Q}}_{\text{in}}^{(k)} \right) ,
\end{eqnarray}
where $\hat{\mathbf{Q}}_{\text{in}}$ and $\hat{\mathbf{Q}}_{\text{loop}}$ are the quadrature operator vector of the input and the cavity pulse (respectively) before the coupling, $R$ is the BS reflectivity and $S$ is the symplectic transformation that the nonlinear medium applies. To find an expression for the observables we compute the expressions of the \textit{covariance matrix}, $\Gamma_{\text{loop}}^{(k)} = \left\langle \hat{\mathbf{Q}}_{\text{loop}}^{(k)} \hat{\mathbf{Q}}_{\text{loop}}^{(k) \top} \right\rangle - \left\langle \hat{\mathbf{Q}}_{\text{loop}}^{(k)}\right\rangle \left\langle \hat{\mathbf{Q}}_{\text{loop}}^{(k)}\right\rangle^{\top}$, where the averages are taken from an ensemble of realizations. As we are working in squeezed vacuum, the covariance matrix and the second-order moments are identical. Also, as we are dealing with Gaussian states, the higher-order moments are functions of the first and second-order moments of the quadratures. It can be shown that the covariance matrix of the cavity pulse at the next time step can be written as
\begin{equation} \label{CM-loop-simple}
    \Gamma_{\text{loop}}^{(k+1)} = R S\Gamma_{\text{loop}}^{(k)}S^{\top} + (1-R) S\Gamma_{\text{in}}^{(k)}S^{\top}  ,
\end{equation}
which can be further expanded by recursion to yield
\begin{equation} \label{CM-loop-expanded}
    \Gamma_{\text{loop}}^{(k+1)} = (1-R) S \left[\sum_{d=0}^{\infty} A^{d} \Gamma_{\text{in}}^{(k-d)} \left(A^{d}\right)^{\top} \right] S^{\top}  ,
\end{equation}
for $A = \sqrt{R} S$. The details of the derivations of Eqs. \eqref{CM-loop-simple} and \eqref{CM-loop-expanded} can be found in our previous work \cite{garciabeni2023}. In Eq. \eqref{CM-loop-expanded}, the index $d$ resembles the delay of the input. 
\subsection{Input memory decay} \label{App-memory-decay}
From Eq. \eqref{CM-loop-expanded} we can see that the magnitude decay of the input information ($\Gamma_{\text{in}}$) depends on the powers of $A$, $A^{d}$, where $d$ denotes the delay of the encoded input. The accessible memory of the reservoir is determined by how much this input information decays each round trip. To quantify this decay we use the spectral norm or L2-norm, defined as
\begin{equation} \label{spectral-norm}
    \lVert A \rVert_{2} = \max_{|\mathbf{x}|\neq0} \frac{|A \mathbf{x}|}{|\mathbf{x}|} \equiv \xi_{\text{max}}(A)  ,
\end{equation}
where $|\cdot|$ stands for the usual euclidian norm of a vector and $\xi_{\text{max}}(A)$ is the maximum singular eigenvalue of $A$ \cite{horn_johnson_1985}. From the definition of $A$, we have that $\lVert A^{d} \rVert_{2} = R^{d/2} \lVert S^{d}\rVert_{2}$. If the cavity is passive, we have that $\lVert S^{d} \rVert_{2} = \lVert S \rVert_{2} = 1$, and so $\lVert A^{d} \rVert_{2} = R^{d/2}$. If the cavity is active we have that $\lVert A^{d} \rVert_{2} \geq R^{d/2}$. For the RC protocol to work, it is important that both the echo state property and the fading memory condition hold \cite{Konkoli2017}. It can be shown that these conditions are fulfilled iff $\rho(A) < 1$, where $\rho(\cdot)$ stands for the spectral radius of a matrix \cite{Nokkala2021}. Moreover, if $\rho(A) < 1$ then the following is also true:
\begin{equation}
    \lim_{d \to \infty} \lVert A^{d} \rVert_{2} = 0  .
\end{equation}
This ensures that the energy of the system does not diverge. We checked that the condition $\rho(A) < 1$ is held in every simulation in the manuscript.

\end{document}